%% file: main.tex
\pdfoutput=1
\documentclass[12pt,a4paper]{article}
\usepackage{ifthen}
\newboolean{pdflatex}
\setboolean{pdflatex}{true}

\newboolean{articletitles}
\setboolean{articletitles}{true} 

\newboolean{uprightparticles}
\setboolean{uprightparticles}{false} 

\newboolean{inbibliography}
\setboolean{inbibliography}{false} 

\input{preamble}
\usepackage{longtable} 

\begin{document}

%%%%%%%%%%%%%%%%%%%%%%%%%
%%%%% Title     %%%%%%%%%
%%%%%%%%%%%%%%%%%%%%%%%%%
\renewcommand{\thefootnote}{\fnsymbol{footnote}}
\setcounter{footnote}{1}

\input{title-LHCb-PAPER}

\renewcommand{\thefootnote}{\arabic{footnote}}
\setcounter{footnote}{0}

%%%%%%%%%%%%%%%%%%%%%%%%%
%%%%% Main text %%%%%%%%%
%%%%%%%%%%%%%%%%%%%%%%%%%

\pagestyle{plain}
\setcounter{page}{1}
\pagenumbering{arabic}

\begin{figure}[!b]
  \begin{center}
    \includegraphics[width=0.45\textwidth]{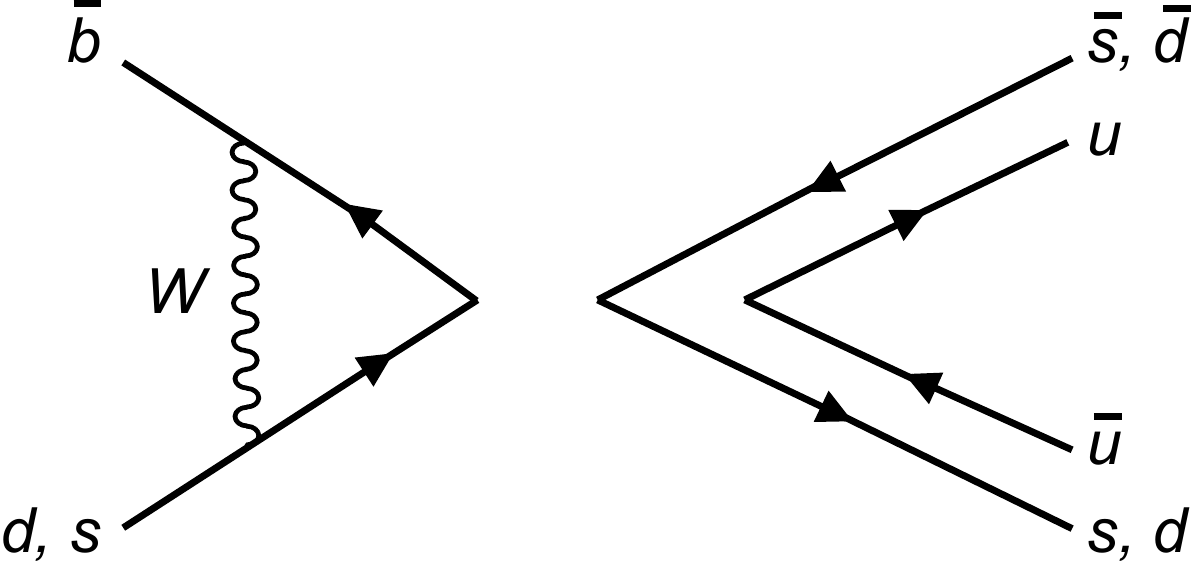}\hspace{0.3cm}
    \includegraphics[width=0.45\textwidth]{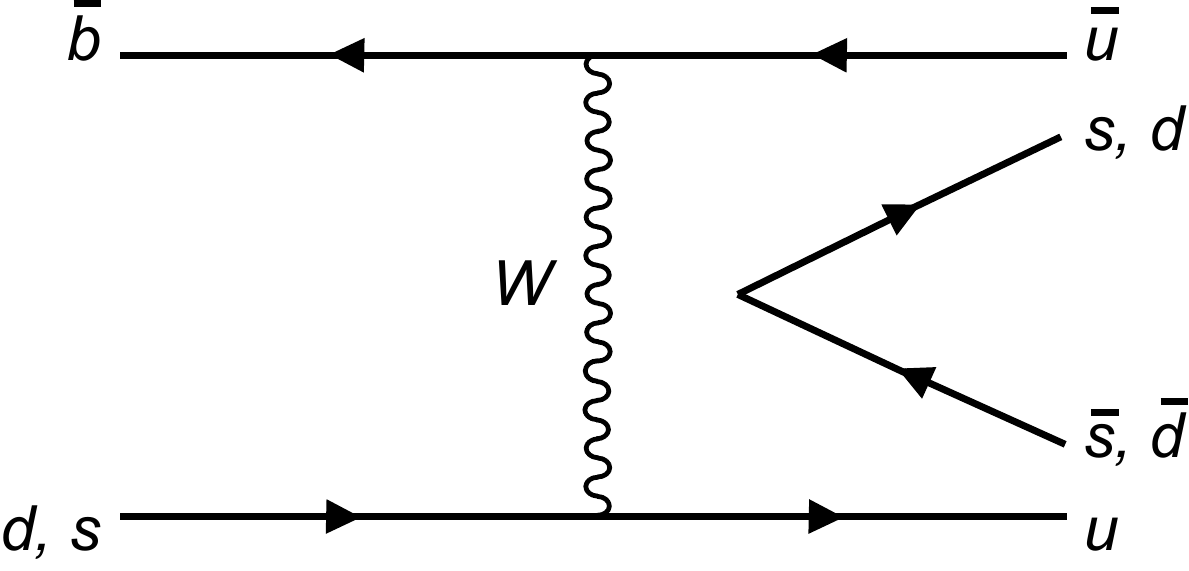}
  \end{center}
  \caption{\small Dominant Feynman graphs contributing to the \BdToKK and \BsTopipi decay amplitudes: (left) penguin-annihilation and (right) $W$-exchange topologies.}
  \label{fig:diagrams}
\end{figure}

The understanding of the dynamics governing the decays of heavy-flavored hadrons is a fundamental ingredient in the search for new particles and new interactions beyond those included in the Standard Model of particle physics (SM). The comparison of theoretical predictions and experimental measurements enables the validity of the SM to be tested up to energy scales well beyond those directly accessible by current particle accelerators. In the last two decades, the development of effective theories significantly improved the accuracy of theoretical predictions for the partial widths of such decays. Several approaches are used to deal with the complexity of quantum chromodynamics (QCD) computations, like QCD factorization (QCDF)~\cite{Beneke:2003zv,Cheng:2009mu,Cheng:2009cn}, perturbative QCD (pQCD)~\cite{Ali:2007ff,Xiao:2011tx} and soft collinear effective theory (SCET)~\cite{Bauer:2004tj}. Despite the general progress in the field, calculations of decay amplitudes governed by so-called weak annihilation transitions are still affected by large uncertainties. In the SM, the rare decay modes \BdToKK and \BsTopipi (charge conjugate modes are implied throughout) can proceed only through such transitions, whose contributions are expected to be small but could be enhanced through certain rescattering effects~\cite{Gronau:2012gs}. The corresponding Feynman graphs are shown in Fig.~\ref{fig:diagrams}. Precise knowledge of the branching fractions of these decays is thus needed to improve our understanding of QCD dynamics in the more general sector of two-body $b$-hadron decays. The $\BdToKK$ and $\BsTopipi$ decays play also a role in techniques proposed to measure the angle $\gamma$ of the unitary triangle~\cite{Fleischer:1999pa}.

While the \BsTopipi decay has already been observed~\cite{LHCb-PAPER-2012-002}, no evidence exists for the \BdToKK decay to date, despite searches performed by the BaBar~\cite{Aubert:2006fha}, CDF~\cite{Aaltonen:2011jv}, Belle~\cite{Duh:2012ie} and LHCb~\cite{LHCb-PAPER-2012-002} collaborations. Averages of the measurements of the branching fractions of these two decays are given by the Heavy Flavor Averaging Group (HFAG): $\BR(\BdToKK)= (0.13^{\,+\,0.06}_{\,-\,0.05}) \times 10^{-6}$ (corresponding to an upper limit of $0.23 \times 10^{-6}$ at 95\% confidence level) and $\BR(\BsTopipi)= (0.76 \pm 0.13) \times 10^{-6}$~\cite{HFAG}. The results of a new search for the \BdToKK decay and an update of the branching fraction measurement of the \BsTopipi decay are presented in this Letter. The data sample that is analyzed corresponds to integrated luminosities of 1.0\invfb at \sqs=\,7\tev and 2.0\invfb at \sqs=\,8\tev of \proton\proton collision data collected with the \lhcb detector in 2011 and 2012, respectively. 

The \lhcb detector~\cite{Alves:2008zz,LHCb-DP-2014-002} is a single-arm forward spectrometer covering the pseudorapidity range $2 < \eta < 5$. The tracking system consists of a silicon-strip vertex detector surrounding the $pp$ interaction region, a large-area silicon-strip detector located upstream of a dipole magnet with a bending power of about
$4{\mathrm{\,Tm}}$, and three stations of silicon-strip detectors and straw drift tubes placed downstream of the magnet. The particle identification (PID) system consists of two ring-imaging Cherenkov (RICH) detectors, scintillating-pad and preshower detectors, electromagnetic
and hadronic calorimeters, and a set of multiwire proportional chambers alternated with iron absorbers.

Simulated events are used in various steps of the analysis. In the simulation, $pp$ collisions are generated using \pythia~\cite{Sjostrand:2006za,Sjostrand:2007gs} with a specific \lhcb configuration~\cite{LHCb-PROC-2010-056}. 
The interaction of the generated particles with the detector, and its response, are implemented using the \geant toolkit~\cite{Allison:2006ve, *Agostinelli:2002hh}, as described in Ref.~\cite{LHCb-PROC-2011-006}.

The online event selection is performed by a trigger~\cite{LHCb-DP-2012-004},
which consists of a hardware stage, based on information from the calorimeter and muon systems, followed by a software stage, which applies a full event reconstruction and requires a secondary vertex (SV) with a significant displacement from all primary \proton\proton interaction vertices (PVs). At least one charged particle must have high transverse momentum, $\pt$, and large $\chisqip$ with respect to all PVs, where \chisqip is the difference between the \chisq of the PV fit performed with and without the considered particle. An algorithm based on a boosted decision tree (BDT) multivariate classifier~\cite{Breiman,Roe} is used for the identification of secondary vertices consistent with the decays of \bquark hadrons~\cite{BBDT}. To further increase the trigger efficiency, an exclusive selection algorithm for two-body $b$-hadron decays was put in place, imposing requirements on the following quantities: the quality of the reconstructed tracks, their \pt and impact parameter (IP), the distance of closest approach between the two oppositely charged tracks used to reconstruct the $b$-hadron candidate, and the \pt, IP and proper decay time of the $b$-hadron candidate.

The event selection is refined offline using another BDT classifier and requirements on PID variables. The BDT returns a discriminant variable which is used to classify each $b$-hadron candidate as either signal or background. 
With the exception of the $b$-hadron decay time, the input variables to 
the BDT classifier are those used in the software trigger, plus the 
following: the largest \pt and IP of the $b$-hadron decay products; the \chisqip of the $b$-hadron candidate; the \chisq of the SV fit; and information on the separation of the SV from the PV. In the presence of multiple PVs per event (up to six and with an average of about two in this analysis), the one with the smallest \chisqip of the $b$-hadron candidate is considered.

The PID system is used to separate the data into mutually exclusive subsamples corresponding to various hypotheses for the final state, namely \Kp\pim, \proton\Km, \proton\pim, as well as \pip\pim and \Kp\Km. The calibration of the PID variables is necessary to determine the yields of other two-body \bquark-hadron decays where one or two particles in the final state are misidentified (cross-feed backgrounds). The efficiencies for a given PID requirement are determined using samples of kaons and pions from the \decay{\Dstarp}{\Dz(\to\Km\pip)\pip} decay chain and protons from \decay{\Lz}{\proton\pim} and \decay{\Lc}{\proton\Km\pip} decays. Since the RICH-based PID information depends on particle momentum, pseudorapidity and track multiplicity, the efficiencies are determined in bins of these variables. They are then averaged over the momentum and pseudorapidity distributions of the final state particles of two-body $b$-hadron decays, and over the distribution of track multiplicity in the corresponding events. Uncertainties on the PID efficiencies are due to the finite sizes of the calibration samples and to the binning used to calculate the efficiencies. The size of the latter uncertainty is estimated by the maximum variation when repeating the PID calibration procedure using different binning schemes. 

The final selection criteria on the BDT output and PID variables are separately optimized for the \BdToKK and \BsTopipi decays. The outcome of the optimization consists of two event selections, $S_{\Kp\Km}$ and $S_{\pip\pim}$, aiming at the best sensitivity on the \BdToKK and \BsTopipi signal yields, respectively. 
In the two selections, common PID requirements are applied to define the subsamples with final-state mass hypotheses other than \Kp\Km and \pip\pim.
The optimization procedure is based on pseudoexperiments generating \Kp\Km and \pip\pim invariant mass distributions. Fits to these distributions are performed with a model identical to that used for the generation. The $B^0_{(s)} \to K^+K^-$ and $B^0_{(s)} \to \pi^+\pi^-$ components are each described by a sum of two Gaussian functions with a common mean to account for mass resolution effects, with parameters determined from data, convolved with a power-law distribution that accounts for final state radiation~(FSR) effects. In particular, the \BsToKK mass shape is deformed due to FSR in the region where the \BdToKK signal is expected. The power-law distribution is derived from analytical quantum electrodynamics (QED) calculations~\cite{Baracchini:2005wp}, and the correctness of the model is checked against simulated events generated by \photos~\cite{Golonka:2005pn}. \photos simulates QED-photon emissions in decays by calculating $\mathcal{O}(\alpha)$ radiative corrections for charged particles using a leading-log collinear approximation. Within the approximation, the program calculates the amount of bremsstrahlung in the decay and modifies the final state according to the decay topology.
The mass distributions of simulated \B candidates, generated with \photos, are well described by fits performed using the mass model developed in this analysis. The fit results are in excellent agreement with the theoretical values of the FSR parameters calculated according to Ref.~\cite{Baracchini:2005wp} for each of the decay modes under study.

The background due to the random association of two oppositely charged tracks (combinatorial background) is modeled with an exponential function. The backgrounds due to the partial reconstruction of multibody $b$-hadron decays are parameterized by means of ARGUS functions~\cite{Albrecht:1994aa} convolved with the same resolution function used for the signals. In the case of partially reconstructed $\decay{\B}{\Kp\pim\PX}$ decays, where \PX stands for one or more missing particles, and the pion is misidentified as a kaon, an incorrect description may alter the determination of the \BdToKK signal yield. Hence the shape of the mass distribution and the size of this contribution to the $K^+K^-$ mass spectrum are determined from data by studying a sample of events selected with tight $K^+\pi^-$ PID requirements, and accounting for the known effects of different PID selection criteria on the invariant mass resolution. The shapes of the mass distributions for cross-feed backgrounds are determined by means of a kernel estimation method~\cite{Cranmer:2001aa} applied to the invariant mass distributions of simulated two-body $b$-hadron decays. As the \BdToKpi cross-feed background contributes to the $K^+K^-$ mass distribution in the \BdToKK signal mass region, the resulting shape of the mass spectrum is validated with data using again a sample of events selected with tight $K^+\pi^-$ PID requirements. The amounts of cross-feed backgrounds are determined relative to the yields of the \BsToKK and \BdTopipi decays, scaled by the branching fractions, PID efficiencies and $b$-quark hadronization probabilities to form \Bd or \Bs mesons~\cite{fsfd}. 

For a given set of BDT and PID selection requirements, pseudoexperiments are generated with yields and model parameters of the backgrounds as determined from data. Signal decays are injected into simulated mass distributions according to different hypotheses for the values of their branching fractions. For each pseudoexperiment, the significance of the signal under study is computed according to Wilks' theorem~\cite{Wilks:1938dza} as $\sqrt{2\ln{(\mathcal{L_{\rm S+B}}/\mathcal{L_{\rm B}})}}$, where $\mathcal{L_{\rm S+B}}$ and $\mathcal{L_{\rm B}}$ are the likelihoods of the nominal fit and of a fit where the yield of the signal is fixed to zero, respectively. As the \BdToKK decay is still not observed and its branching fraction not well constrained, a multidimensional scan is performed over a wide range of branching fraction values, as well as BDT and PID selection requirements. For each point of the scan the signal significance is determined. The point corresponding to the smallest branching fraction that can be measured with a significance of 5 standard deviations is determined, and the optimal selection requirements are thus identified. This branching fraction is found to be $\BR_{\rm min} \simeq 6 \times 10^{-8}$. In contrast, the expected yield of \BsTopipi decays is more precisely constrained, and the optimization of the selection requirements is found not to depend on the assumed branching fractions within $\pm2$ standard deviations from the current world average value~\cite{HFAG}. The optimization procedure for $S_{\Kp\Km}$ leads to tighter PID and looser BDT requirements with respect to $S_{\pip\pim}$. This is due to the fact that the random association of two kaons is much less likely than that of two pions, and thus the correct identification of two kaons provides a more powerful rejection of the combinatorial background with respect to that of two pions. As a consequence, the combinatorial background in the \pip\pim spectrum is best suppressed by the application of tighter requirements on the BDT output.

\begin{figure*}[!t]
  \begin{center}
    \includegraphics[width=0.48\textwidth]{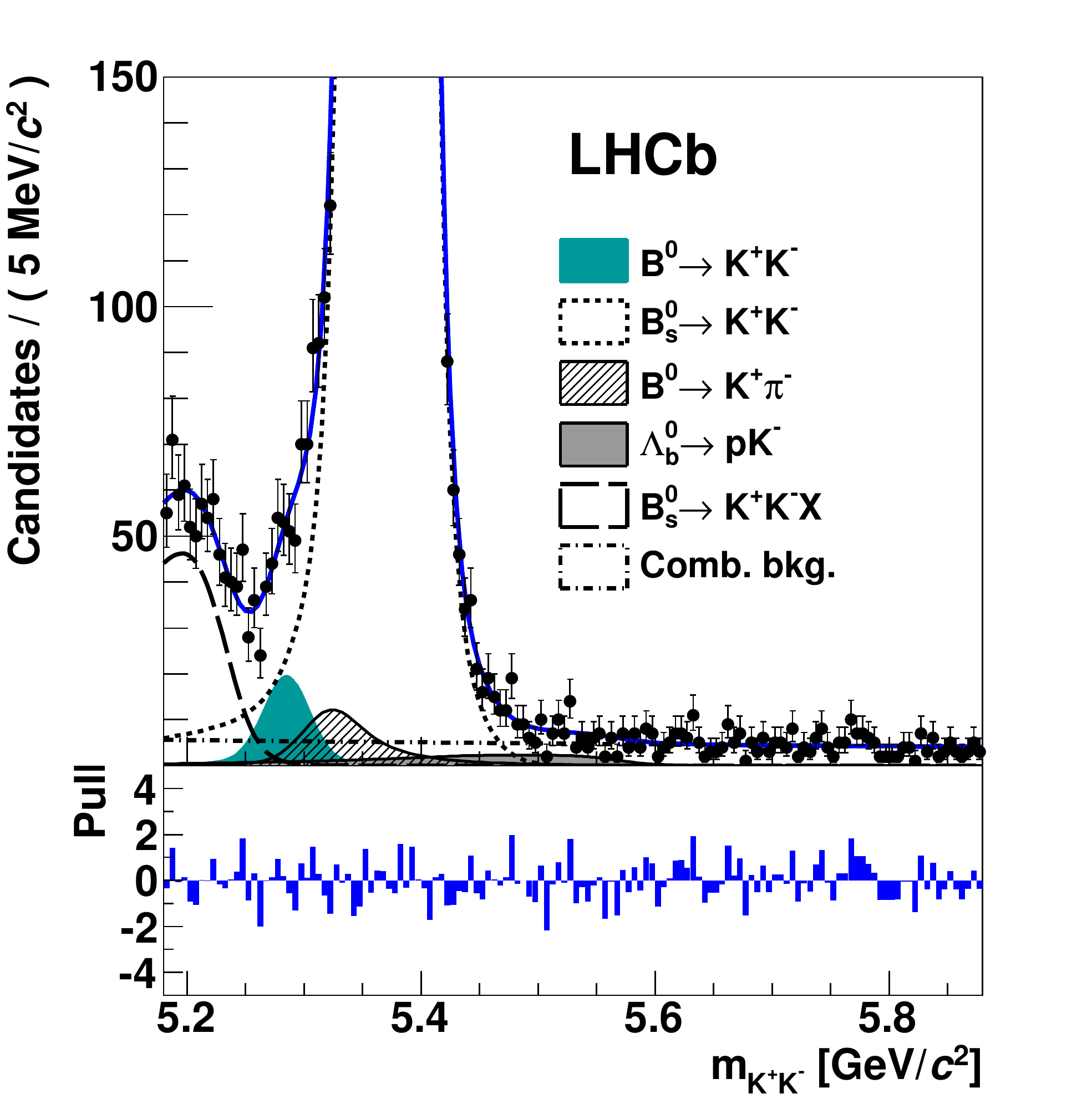}
    \includegraphics[width=0.48\textwidth]{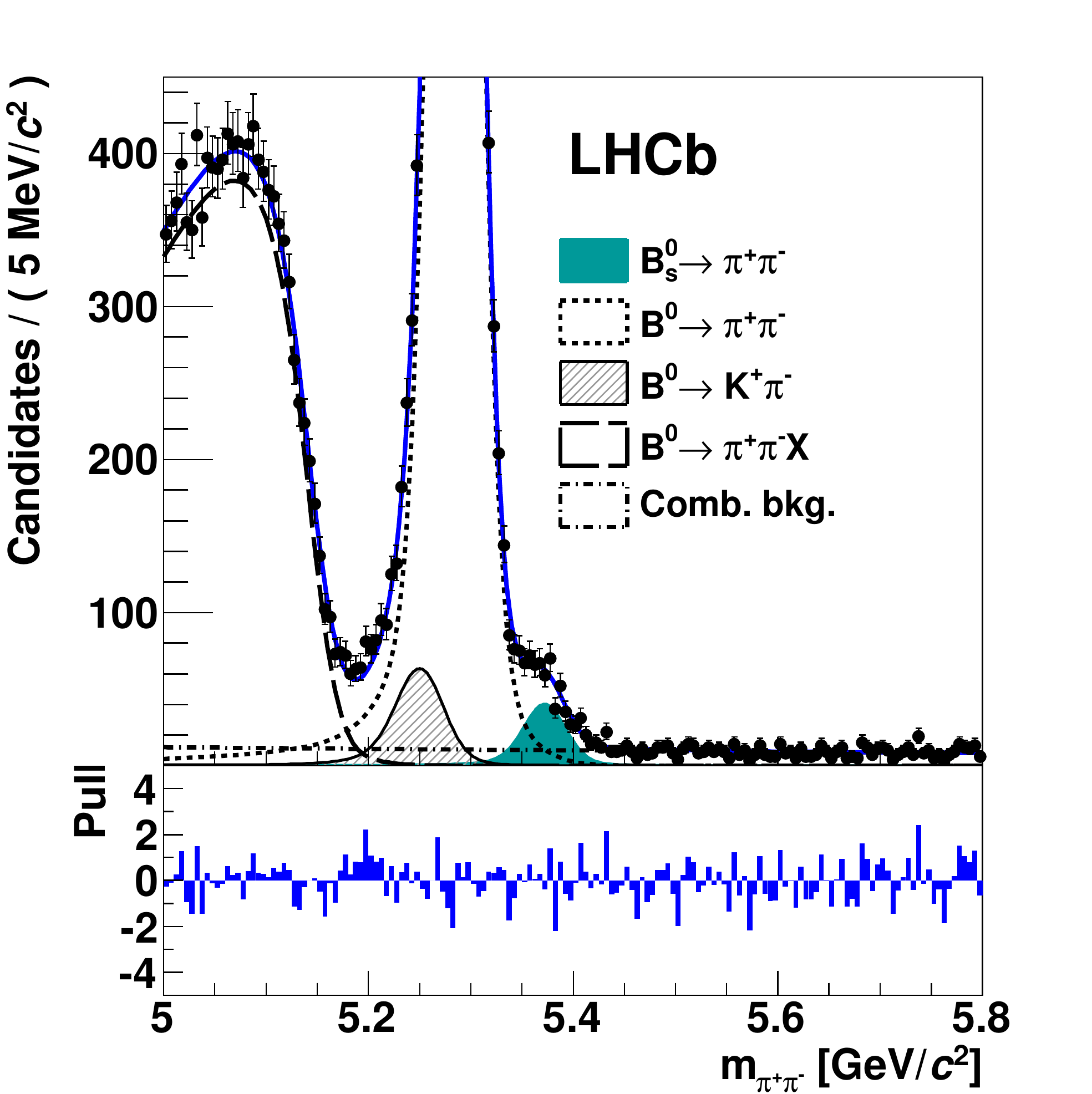}
  \end{center}
  \caption{\small Distributions of (left) $m_{\Kp\Km}$ and (right) $m_{\pip\pim}$ for candidates passing $S_{\Kp\Km}$ and $S_{\pip\pim}$, respectively. The continuous (blue) curves represent the results of the best fits to the data points. The most relevant contributions to the invariant mass spectra are shown as indicated in the legends. The vertical scales are chosen to magnify the relevant signal regions. The bin-by-bin differences between the fits and the data, in units of standard deviations, are also shown.}
  \label{fig:massPlots}
\end{figure*}

After applying the BDT and PID criteria for $S_{\Kp\Km}$ or $S_{\pip\pim}$, the signal yields are determined by means of an extended binned maximum likelihood fit done simultaneously with the exclusive data sets defined by the different mass hypotheses of particles in the final state. The model fitted to the mass distributions is the same as that used in the optimization of the selection. The amount of each cross-feed background contribution is determined directly from the fits, taking into account the appropriate PID efficiency factors. The $m_{\Kp\Km}$ and $m_{\pip\pim}$ invariant mass distributions are shown in Fig.~\ref{fig:massPlots}, with the results of the best fits superimposed. 
The yields for the two signals are $N(\BdToKK) = 201 \pm 33 \pm 14$ and $N(\BsTopipi) = 455 \pm 35 \pm 24$, where the first uncertainty is statistical and the second is systematic. The systematic uncertainties are related to the choice of the model used to parameterize the invariant mass shapes of signal and background components and to the knowledge of the PID efficiencies used to determine the amount of cross-feed backgrounds. The results of the best fits are used to generate pseudoexperiments, and then fits with alternative models are applied to the mass distributions. By studying the distributions of the difference between the signal yields determined from the nominal fit and those performed with alternative models, systematic uncertainties are determined. Such alternative models are considered for signal, combinatorial background, background from partially reconstructed \bquark-hadron decays and cross-feed background mass models. The systematic uncertainty due to PID efficiencies is also assessed by generating pseudoexperiments and fitting the nominal model to the output mass distributions, using PID efficiencies randomly varied in each pseudoexperiment according to their estimated uncertainties. The standard deviation of the distribution of the yields determined in each set of pseudoexperiments is taken as a systematic uncertainty. The contributions of the various systematic uncertainties are reported in Table~\ref{tab:systematics}. The systematic uncertainties associated to the knowledge of the cross-feed background mass shapes are found to be negligible and are not reported. The total systematic uncertainties are obtained by summing all contributions in quadrature.

\begin{table}[!b]
  \caption{ \small Systematic uncertainties on the yields for the \BdToKK and \BsTopipi decays.}
  \begin{center}
    \begin{tabular}{lcc}
      Systematic uncertainty     & $N(\BdToKK)$ & $N(\BsTopipi)$ \\
      \hline 
      Signal mass shape              & $11.8$      & $\phantom{2}6.3$ \\
      Combinatorial mass shape   & $\phantom{1}5.5$       & $\phantom{2}2.6$ \\
      Partially reco. mass shape   & $\phantom{1}1.3$       & $23.1$ \\
      PID efficiencies                   & $\phantom{1}3.4$       & $\phantom{2}2.5$ \\
      \hline
      Sum in quadrature             & $13.5$      & $24.2$ \\
    \end{tabular} 
  \end{center}
  \label{tab:systematics}
\end{table}

\begin{figure}[!t]
  \begin{center}
    \includegraphics[width=\columnwidth]{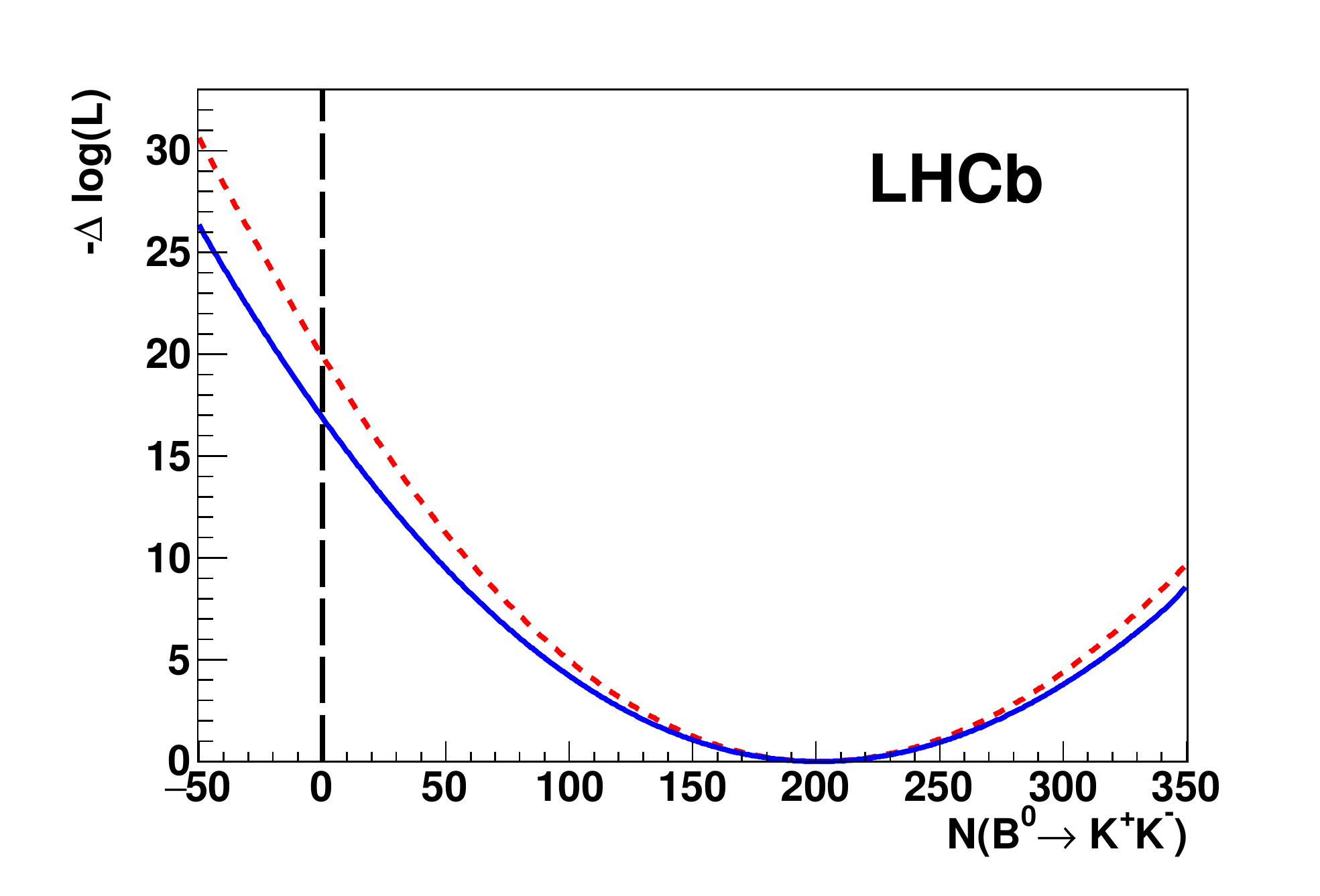}
  \end{center}
  \caption{\small Log-likelihood ratio as a function of the \BdToKK signal yield. The dashed (red) and continuous (blue) curves correspond to the exclusion and to the inclusion of systematic uncertainties, respectively.}
  \label{fig:significance}
\end{figure}

The significance of the \BdToKK signal with respect to the null hypothesis is determined by means of a profile likelihood ratio. To account for systematic uncertainties, the likelihood function is convolved with a Gaussian function with width equal to the systematic uncertainty. The log-likelihood ratio as a function of the \BdToKK signal yield is shown in Fig.~\ref{fig:significance}. The statistical significance is found to be 6.3 standard deviations, reduced to 5.8 when considering systematic uncertainties.

The branching fractions of \BdToKK and \BsTopipi decays are determined relative to the \BdToKpi branching fraction, according to the following equation
\begin{equation}\label{eq:relativeBR}
	\frac{f_x}{f_d}\,\frac{\BR(\decay{\B^0_x}{\Ph^+\Ph^-})}{\BR(\BdToKpi)} =
	\frac{N(\decay{\B^0_x}{\Ph^+\Ph^-})}{N(\BdToKpi)}\,\frac{\varepsilon(\BdToKpi)}{\varepsilon(\decay{\B^0_x}{\Ph^+\Ph^-})},\nonumber
\end{equation}
where $f_x$ is the probability for a \bquark quark to hadronize into a $\B^0_x$ meson ($x=d,s$), $N$ and $\varepsilon$ are the yield and the efficiency for the given decay mode, respectively, and $h$ stands for $K$ or $\pi$. 
The yields of the \BdToKpi decay in the sub-samples selected with \Kp\pim PID requirements are determined from the fits, and their values are $N(\BdToKpi) = 105\hspace{0.5mm}010 \pm 431 \pm 988$ and $N(\BdToKpi) = 71\hspace{0.5mm}304 \pm 312 \pm 609$, when applying the BDT requirements of $S_{\Kp\Km}$ and $S_{\pip\pim}$, respectively.
Trigger and reconstruction efficiencies are determined from simulation and corrected using information from data. 
For the \BsTopipi decay the sizeable value of the decay width difference between the long- and short-lived components of the \Bs-meson system is taken into account. The \BsTopipi lifetime is assumed to be that of the short-lived component, as expected in presence of small \CP violation. The final ratios of efficiencies are found to be $2.08 \pm 0.16$ and $1.43 \pm 0.10$ for the \BdToKK and \BsTopipi decays, respectively. The dominant contributions to the uncertainties on these ratios are due to the PID calibration and to the knowledge of the trigger efficiencies. The following results are then obtained
\begin{eqnarray}
	\frac{\BR(\BdToKK)}{\BR(\BdToKpi)} & = & (3.98 \pm 0.65 \pm 0.42) \times 10^{-3}, \nonumber \\
	\frac{f_s}{f_d}\,\frac{\BR(\BsTopipi)}{\BR(\BdToKpi)} & = & (9.15 \pm 0.71 \pm 0.83) \times 10^{-3}, \nonumber
\end{eqnarray}
where the first uncertainty is statistical and the second systematic. Using the HFAG average $\BR(\BdToKpi) = (19.57^{+0.53}_{-0.52}) \times 10^{-6}$~\cite{HFAG}, and $f_s/f_d = 0.259 \pm 0.015$ from Ref.~\cite{fsfd}, the following branching fractions are obtained
{\small
\begin{eqnarray}
	\BR(\BdToKK) & = & (7.80 \pm 1.27 \pm 0.81 \pm 0.21) \times 10^{-8}, \nonumber \\
	\BR(\BsTopipi) & = & (6.91 \pm 0.54 \pm 0.63 \pm 0.19 \pm 0.40) \times 10^{-7},\nonumber
\end{eqnarray}
}where the first uncertainty is statistical, the second systematic, and the third and fourth are due to the knowledge of $\BR(\BdToKpi)$ and of $f_s / f_d$, respectively.

Various theoretical predictions of the branching fractions of \BdToKK and \mbox{\BsTopipi} decays are available in the literature~\cite{Lu:2005mx,Ali:2007ff,Cheng:2009mu,Cheng:2009cn,Xiao:2011tx,Gronau:2012gs,Wang:2013fya,Chang:2014rla,Chang:2014yma,Li:2015xna}. The pQCD estimations in Ref.~\cite{Xiao:2011tx} are in agreement within uncertainties with the present results. The QCDF prediction of $\BR(\BdToKK)$ in Ref.~\cite{Cheng:2009mu} agrees well with these results, but that of $\BR(\BsTopipi)$ is significantly smaller than the measurement. In Ref.~\cite{Chang:2014yma}, the unexpectedly large value of $\BR(\BsTopipi)$ caused the traditional QCDF treatment for annihilation parameters to be revisited. 

In summary, this Letter reports the most precise measurements of the branching fractions for the \BdToKK and \BsTopipi decay modes to date. These are in good agreement with and supersede those reported in Ref.~\cite{LHCb-PAPER-2012-002}, which were the best results available prior to the present analysis. The \BdToKK decay is the rarest fully hadronic $B$-meson decay ever observed.

\input{acknowledgements}

\addcontentsline{toc}{section}{References}
\setboolean{inbibliography}{true}
\bibliographystyle{LHCb}
\bibliography{main}

\newpage
\input{LHCb_Authorship_flat_13-Jul-2016}

\end{document}

%% file: preamble.tex
\usepackage[top=1in, bottom=1.25in, left=1in, right=1in]{geometry}

\usepackage{microtype}
\usepackage{lineno} 
\usepackage{xspace} 
\usepackage{caption}

\usepackage{graphicx}  % to include figures (can also use other packages)
\usepackage{color}
\usepackage{colortbl}
\graphicspath{{./figs/}} % Make Latex search fig subdir for figures

\usepackage{amsmath} % Adds a large collection of math symbols
\usepackage{amssymb}
\usepackage{amsfonts}
\usepackage{upgreek} % Adds in support for greek letters in roman typeset

\newcommand*\patchAmsMathEnvironmentForLineno[1]{%
\expandafter\let\csname old#1\expandafter\endcsname\csname #1\endcsname
\expandafter\let\csname oldend#1\expandafter\endcsname\csname
end#1\endcsname
 \renewenvironment{#1}%
   {\linenomath\csname old#1\endcsname}%
   {\csname oldend#1\endcsname\endlinenomath}%
}
\newcommand*\patchBothAmsMathEnvironmentsForLineno[1]{%
  \patchAmsMathEnvironmentForLineno{#1}%
  \patchAmsMathEnvironmentForLineno{#1*}%
}
\AtBeginDocument{%
\patchBothAmsMathEnvironmentsForLineno{equation}%
\patchBothAmsMathEnvironmentsForLineno{align}%
\patchBothAmsMathEnvironmentsForLineno{flalign}%
\patchBothAmsMathEnvironmentsForLineno{alignat}%
\patchBothAmsMathEnvironmentsForLineno{gather}%
\patchBothAmsMathEnvironmentsForLineno{multline}%
\patchBothAmsMathEnvironmentsForLineno{eqnarray}%
}

\usepackage{hyperref}    % Hyperlinks in references
\usepackage[all]{hypcap} % Internal hyperlinks to floats.

\input{lhcb-symbols-def} % Add in the predefined LHCb symbols

\usepackage{cite} % Allows for ranges in citations
\usepackage{mciteplus}

%% file: lhcb-symbols-def.tex
\usepackage{xspace} 
\usepackage{upgreek}

\def\lhcb {\mbox{LHCb}\xspace}

\def\MagUp {\mbox{\em Mag\kern -0.05em Up}\xspace}

\ifthenelse{\boolean{uprightparticles}}%
{

 \def\Ppi         {\ensuremath{\uppi}\xspace}

 \def\PDelta      {\ensuremath{\Delta}\xspace}                 
 \def\PXi      {\ensuremath{\Xi}\xspace}                 
 \def\PLambda      {\ensuremath{\Lambda}\xspace}                 
 \def\PSigma      {\ensuremath{\Sigma}\xspace}                 
 \def\POmega      {\ensuremath{\Omega}\xspace}                 
 \def\PUpsilon      {\ensuremath{\Upsilon}\xspace}

 \def\PB      {\ensuremath{\mathrm{B}}\xspace}                 
                  
 \def\PD      {\ensuremath{\mathrm{D}}\xspace}

 \def\PK      {\ensuremath{\mathrm{K}}\xspace}

 \def\PX      {\ensuremath{\mathrm{X}}\xspace}

 \def\Pb      {\ensuremath{\mathrm{b}}\xspace}                 
 \def\Pc      {\ensuremath{\mathrm{c}}\xspace}

 \def\Ph      {\ensuremath{\mathrm{h}}\xspace}                 
 \def\Pi      {\ensuremath{\mathrm{i}}\xspace}

 \def\Pp      {\ensuremath{\mathrm{p}}\xspace}

 \def\Ps      {\ensuremath{\mathrm{s}}\xspace}

}
{

 \def\Ppi         {\ensuremath{\pi}\xspace}

 \mathchardef\PDelta="7101
 \mathchardef\PXi="7104
 \mathchardef\PLambda="7103
 \mathchardef\PSigma="7106
 \mathchardef\POmega="710A
 \mathchardef\PUpsilon="7107
                  
 \def\PB      {\ensuremath{B}\xspace}                 
                  
 \def\PD      {\ensuremath{D}\xspace}

 \def\PK      {\ensuremath{K}\xspace}

 \def\PX      {\ensuremath{X}\xspace}

 \def\Pb      {\ensuremath{b}\xspace}                 
 \def\Pc      {\ensuremath{c}\xspace}

 \def\Ph      {\ensuremath{h}\xspace}                 
 \def\Pi      {\ensuremath{i}\xspace}

 \def\Pp      {\ensuremath{p}\xspace}

 \def\Ps      {\ensuremath{s}\xspace}

}

\makeatletter
\ifcase \@ptsize \relax% 10pt
  \newcommand{\miniscule}{\@setfontsize\miniscule{4}{5}}% \tiny: 5/6
\or
  \newcommand{\miniscule}{\@setfontsize\miniscule{5}{6}}% \tiny: 6/7
\or
  \newcommand{\miniscule}{\@setfontsize\miniscule{5}{6}}% \tiny: 6/7
\fi
\makeatother

\DeclareRobustCommand{\optbar}[1]{\shortstack{{\miniscule (\rule[.5ex]{1.25em}{.18mm})}
  \\ [-.7ex] $#1$}}

\def\squark    {{\ensuremath{\Ps}}\xspace}

\def\cquark    {{\ensuremath{\Pc}}\xspace}

\def\bquark    {{\ensuremath{\Pb}}\xspace}

\def\pion   {{\ensuremath{\Ppi}}\xspace}

\def\pip    {{\ensuremath{\pion^+}}\xspace}
\def\pim    {{\ensuremath{\pion^-}}\xspace}

\def\kaon    {{\ensuremath{\PK}}\xspace}
  \def\Kbar    {{\kern 0.2em\overline{\kern -0.2em \PK}{}}\xspace}

\def\KorKbar    {\kern 0.18em\optbar{\kern -0.18em K}{}\xspace}

\def\Kp      {{\ensuremath{\kaon^+}}\xspace}
\def\Km      {{\ensuremath{\kaon^-}}\xspace}

  \def\Dbar    {{\kern 0.2em\overline{\kern -0.2em \PD}{}}\xspace}
\def\D       {{\ensuremath{\PD}}\xspace}

\def\DorDbar    {\kern 0.18em\optbar{\kern -0.18em D}{}\xspace}
\def\Dz      {{\ensuremath{\D^0}}\xspace}

\def\Dstarp  {{\ensuremath{\D^{*+}}}\xspace}

\def\B       {{\ensuremath{\PB}}\xspace}
\def\Bbar    {{\ensuremath{\kern 0.18em\overline{\kern -0.18em \PB}{}}}\xspace}

\def\BorBbar    {\kern 0.18em\optbar{\kern -0.18em B}{}\xspace}

\def\Bd      {{\ensuremath{\B^0}}\xspace}
\def\Bs      {{\ensuremath{\B^0_\squark}}\xspace}

  \def\Y#1S{\ensuremath{\PUpsilon{(#1S)}}\xspace}% no space before {...}!

\def\proton      {{\ensuremath{\Pp}}\xspace}

\def\Lz          {{\ensuremath{\PLambda}}\xspace}
\def\Lbar        {{\ensuremath{\kern 0.1em\overline{\kern -0.1em\PLambda}}}\xspace}
\def\LorLbar    {\kern 0.18em\optbar{\kern -0.18em \PLambda}{}\xspace}

\def\Lc      {{\ensuremath{\Lz^+_\cquark}}\xspace}

\def\BF         {{\ensuremath{\mathcal{B}}}\xspace}

\def\BR         {\BF}
\newcommand{\decay}[2]{\ensuremath{#1\!\to #2}\xspace}         % {\Pa}{\Pb \Pc}

\def\to                 {\ensuremath{\rightarrow}\xspace}

\def\CP                {{\ensuremath{C\!P}}\xspace}

\def\BdTopipi     {\decay{\Bd}{\pip\pim}}
\def\BsTopipi     {\decay{\Bs}{\pip\pim}}
\def\BdToKpi      {\decay{\Bd}{\Kp\pim}}
\def\BsToKK       {\decay{\Bs}{\Kp\Km}}
\def\BdToKK       {\decay{\Bd}{\Kp\Km}}

\def\AT#1     {\ensuremath{A_{\mathrm{T}}^{#1}}\xspace}           % 2

\def\C#1      {\ensuremath{\mathcal{C}_{#1}}\xspace}                       % 9
\def\Cp#1     {\ensuremath{\mathcal{C}_{#1}^{'}}\xspace}                    % 7
\def\Ceff#1   {\ensuremath{\mathcal{C}_{#1}^{\mathrm{(eff)}}}\xspace}        % 9  
\def\Cpeff#1  {\ensuremath{\mathcal{C}_{#1}^{'\mathrm{(eff)}}}\xspace}       % 7
\def\Ope#1    {\ensuremath{\mathcal{O}_{#1}}\xspace}                       % 2
\def\Opep#1   {\ensuremath{\mathcal{O}_{#1}^{'}}\xspace}                    % 7

\newcommand{\tev}{\ifthenelse{\boolean{inbibliography}}{\ensuremath{~T\kern -0.05em eV}\xspace}{\ensuremath{\mathrm{\,Te\kern -0.1em V}}}\xspace}
\newcommand{\gev}{\ensuremath{\mathrm{\,Ge\kern -0.1em V}}\xspace}
\newcommand{\mev}{\ensuremath{\mathrm{\,Me\kern -0.1em V}}\xspace}
\newcommand{\kev}{\ensuremath{\mathrm{\,ke\kern -0.1em V}}\xspace}
\newcommand{\ev}{\ensuremath{\mathrm{\,e\kern -0.1em V}}\xspace}
\newcommand{\gevc}{\ensuremath{{\mathrm{\,Ge\kern -0.1em V\!/}c}}\xspace}
\newcommand{\mevc}{\ensuremath{{\mathrm{\,Me\kern -0.1em V\!/}c}}\xspace}
\newcommand{\gevcc}{\ensuremath{{\mathrm{\,Ge\kern -0.1em V\!/}c^2}}\xspace}
\newcommand{\gevgevcccc}{\ensuremath{{\mathrm{\,Ge\kern -0.1em V^2\!/}c^4}}\xspace}
\newcommand{\mevcc}{\ensuremath{{\mathrm{\,Me\kern -0.1em V\!/}c^2}}\xspace}

\def\invfb   {\ensuremath{\mbox{\,fb}^{-1}}\xspace}

\newcommand{\chisq}{\ensuremath{\chi^2}\xspace}

\newcommand{\chisqip}{\ensuremath{\chi^2_{\text{IP}}}\xspace}

\def\gsim{{~\raise.15em\hbox{$>$}\kern-.85em
          \lower.35em\hbox{$\sim$}~}\xspace}
\def\lsim{{~\raise.15em\hbox{$<$}\kern-.85em
          \lower.35em\hbox{$\sim$}~}\xspace}

\def\sqs   {\ensuremath{\protect\sqrt{s}}\xspace}

\def\pt         {\mbox{$p_{\mathrm{ T}}$}\xspace}

\def\geant      {\mbox{\textsc{Geant4}}\xspace}

\def\photos     {\mbox{\textsc{Photos}}\xspace}

\def\pythia     {\mbox{\textsc{Pythia}}\xspace}

\def\tell1  {TELL1\xspace}
\def\ukl1   {UKL1\xspace}

%% file: title-LHCb-PAPER.tex
%%%%%%%%%%%%%%%%%%%%%%%%%
%%%%%  TITLE PAGE  %%%%%%
%%%%%%%%%%%%%%%%%%%%%%%%%
\begin{titlepage}
\pagenumbering{roman}

% Header ---------------------------------------------------
\vspace*{-1.5cm}
\centerline{\large EUROPEAN ORGANIZATION FOR NUCLEAR RESEARCH (CERN)}
\vspace*{1.5cm}
\noindent
\begin{tabular*}{\linewidth}{lc@{\extracolsep{\fill}}r@{\extracolsep{0pt}}}
\ifthenelse{\boolean{pdflatex}}
{\vspace*{-2.7cm}\mbox{\!\!\!\includegraphics[width=.14\textwidth]{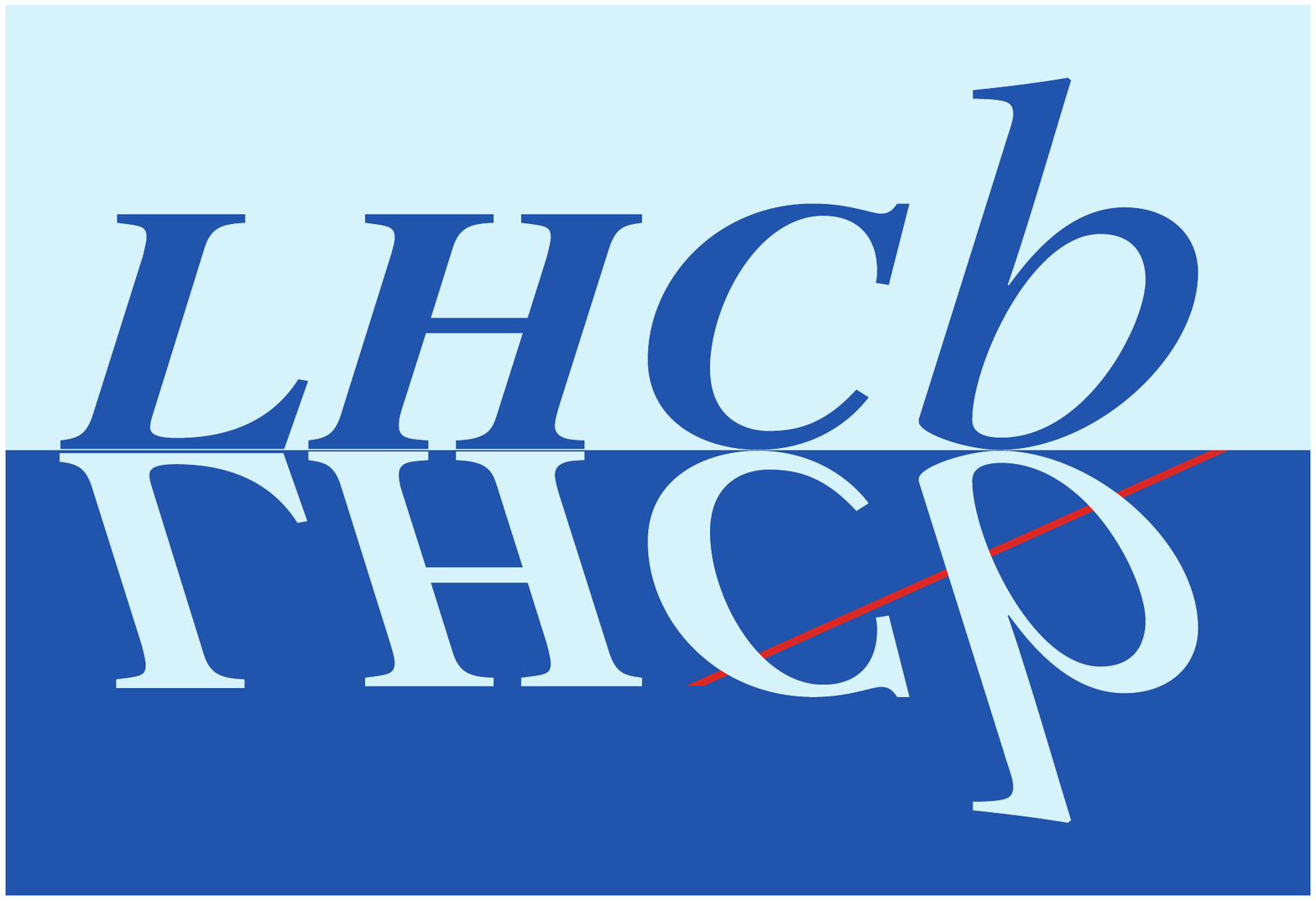}} & &}%
{\vspace*{-1.2cm}\mbox{\!\!\!\includegraphics[width=.12\textwidth]{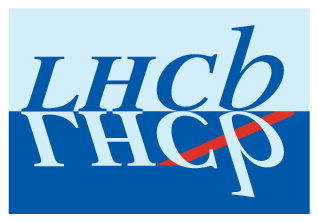}} & &}%
\\
 & & CERN-EP-2016-255 \\  
 & & LHCb-PAPER-2016-036 \\  
 & & February 27, 2017 \\ 
 & & \\

\end{tabular*}

\vspace*{4.0cm}

% Title --------------------------------------------------
{\normalfont\bfseries\boldmath\huge
\begin{center}
Observation of the annihilation decay mode \BdToKK
\end{center}
}

\vspace*{2.0cm}

% Authors -------------------------------------------------
\begin{center}
The LHCb collaboration\footnote{Authors are listed at the end of this letter.}
\end{center}

\vspace{\fill}

% Abstract -----------------------------------------------
\begin{abstract}
  \noindent
\noindent A search for the \BdToKK decay is performed using $pp$-collision data collected by \lhcb. The data set corresponds to integrated luminosities of 1.0 and 2.0\invfb at center-of-mass energies of 7 and 8\tev, respectively. This decay is observed for the first time, with a significance of more than five standard deviations. The analysis also results in an improved measurement of the branching fraction for the \BsTopipi decay. The measured branching fractions are $\BR(\BdToKK) = (7.80 \pm 1.27 \pm 0.81 \pm 0.21) \times 10^{-8}$ and $\BR(\BsTopipi) = (6.91 \pm 0.54 \pm 0.63 \pm 0.19 \pm 0.40) \times 10^{-7}$. The first uncertainty is statistical, the second is systematic, the third is due to the uncertainty on the \BdToKpi branching fraction used as a normalization. For the \Bs mode, the fourth accounts for the uncertainty on the ratio of the probabilities for \bquark quarks to hadronize into \Bs and \Bd mesons.
\end{abstract}

\vspace*{2.0cm}

\begin{center}
  Published in Phys. Rev. Lett. 118 (2017) 081801 
\end{center}

\vspace{\fill}

{\footnotesize 
\centerline{\copyright~CERN on behalf of the \lhcb collaboration, licence \href{http://creativecommons.org/licenses/by/4.0/}{CC-BY-4.0}.}}
\vspace*{2mm}

\end{titlepage}

\newpage
\setcounter{page}{2}
\mbox{~}

\cleardoublepage

%% file: acknowledgements.tex
\section*{Acknowledgements}
\noindent We express our gratitude to our colleagues in the CERN
accelerator departments for the excellent performance of the LHC. We
thank the technical and administrative staff at the LHCb
institutes. We acknowledge support from CERN and from the national
agencies: CAPES, CNPq, FAPERJ and FINEP (Brazil); NSFC (China);
CNRS/IN2P3 (France); BMBF, DFG and MPG (Germany); INFN (Italy); 
FOM and NWO (The Netherlands); MNiSW and NCN (Poland); MEN/IFA (Romania); 
MinES and FASO (Russia); MinECo (Spain); SNSF and SER (Switzerland); 
NASU (Ukraine); STFC (United Kingdom); NSF (USA).
We acknowledge the computing resources that are provided by CERN, IN2P3 (France), KIT and DESY (Germany), INFN (Italy), SURF (The Netherlands), PIC (Spain), GridPP (United Kingdom), RRCKI and Yandex LLC (Russia), CSCS (Switzerland), IFIN-HH (Romania), CBPF (Brazil), PL-GRID (Poland) and OSC (USA). We are indebted to the communities behind the multiple open 
source software packages on which we depend.
Individual groups or members have received support from AvH Foundation (Germany),
EPLANET, Marie Sk\l{}odowska-Curie Actions and ERC (European Union), 
Conseil G\'{e}n\'{e}ral de Haute-Savoie, Labex ENIGMASS and OCEVU, 
R\'{e}gion Auvergne (France), RFBR and Yandex LLC (Russia), GVA, XuntaGal and GENCAT (Spain), Herchel Smith Fund, The Royal Society, Royal Commission for the Exhibition of 1851 and the Leverhulme Trust (United Kingdom).

%% file: LHCb_Authorship_flat_13-Jul-2016.tex
\centerline{\large\bf LHCb collaboration}
\begin{flushleft}
\small
R.~Aaij$^{40}$,
B.~Adeva$^{39}$,
M.~Adinolfi$^{48}$,
Z.~Ajaltouni$^{5}$,
S.~Akar$^{6}$,
J.~Albrecht$^{10}$,
F.~Alessio$^{40}$,
M.~Alexander$^{53}$,
S.~Ali$^{43}$,
G.~Alkhazov$^{31}$
% ...

{\footnotesize \it
% ...
$ ^{5}$Clermont Universit{\'e}, Universit{\'e} Blaise Pascal, CNRS/IN2P3, LPC, Clermont-Ferrand, France\\
$ ^{6}$CPPM, Aix-Marseille Universit{\'e}, CNRS/IN2P3, Marseille, France\\
% ...
$ ^{10}$Fakult{\"a}t Physik, Technische Universit{\"a}t Dortmund, Dortmund, Germany\\
% ...
$ ^{31}$Petersburg Nuclear Physics Institute (PNPI), Gatchina, Russia\\
% ...
$ ^{39}$Universidad de Santiago de Compostela, Santiago de Compostela, Spain\\
$ ^{40}$European Organization for Nuclear Research (CERN), Geneva, Switzerland\\
% ...
$ ^{43}$Nikhef National Institute for Subatomic Physics, Amsterdam, The Netherlands\\
% ...
$ ^{48}$H.H. Wills Physics Laboratory, University of Bristol, Bristol, United Kingdom\\
% ...
$ ^{53}$School of Physics and Astronomy, University of Glasgow, Glasgow, United Kingdom\\
% ...
}
\end{flushleft}